\documentclass[preprint2]{aastex}
\usepackage{graphicx}
\usepackage{color}

\begin{document}


\title{Earth as a potential source of life for Europa’s subsurface ocean}

\author{Z.N. Osmanov}
\affil{School of Physics, Free University of Tbilisi, 0183, Tbilisi,
Georgia}

\affil{E. Kharadze Georgian National Astrophysical Observatory, Abastumani 0301, Georgia}

\begin{abstract}
The paper discusses the possibility of dust particles containing living bacteria ejected from Earth reaching Europa and landing on its surface. It is shown that, taking certain factors into account, over a period of $30–80$ Myr (the estimated age of Europa’s ocean), Jupiter’s moon would have been impacted by approximately $3\times 10^{23}$ to $8\times 10^{23}$ particles in total, within which a bacterium could have survived. In the paper, we discuss the possibility of dust grains entering liquid water beneath the surface.

\end{abstract}


\section{Introduction}

A recent paper by \cite{pansp1} investigates the possibility of panspermia originating from Earth. It has been shown that turbulent motion caused by extraterrestrial micrometeorites entering the atmosphere can involve local dust particles, which may subsequently escape the Solar System and potentially reach up to $10^5$ stellar systems.

In general, the idea of life being transported to Earth from deep space is not new; it was first considered in the early 20th century by \cite{arrhenius}, who argued that solar radiation could propel small cosmic dust particles across vast interstellar distances. The problem has been studied in detail in the context of survival under extraterrestrial conditions \citep{UV,wesson}. A study by \cite{chandra} argues that life on Earth may have an extraterrestrial origin.

Our recent work \citep{pansp1} is based on the results published by \cite{berera} whose study focuses on the dynamics of space dust interactions with atmospheric flow during high-velocity collisions. During these collisions, the local dust particles originating from the Earth might acquire velocities exceeding the Earth's escape velocity. Consequently, bacteria-bearing dust particles may be transported to distant planetary systems.

Life on Earth originated at least $3.55$ billion years ago \citep{life}, which implies that for approximately that long, Earth has been shedding life-bearing particles into surrounding space. Hence, if favorable conditions exist elsewhere in the Solar System and can be accessed by dust particles, the transport of life from Earth appears plausible and may have been occurring over the course of several billion years.

\begin{figure}
  \centering {\includegraphics[width=7cm]{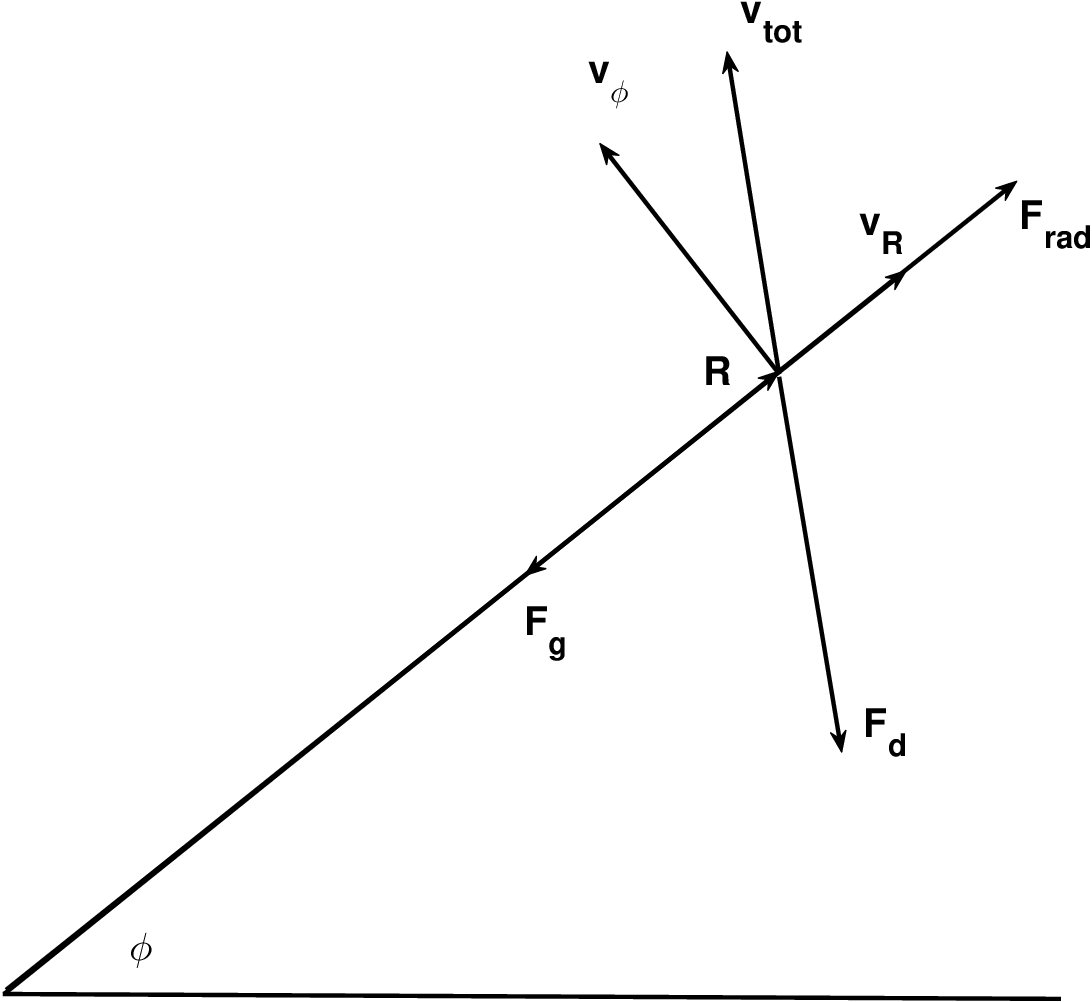}}
  \caption{The schematic picture of the components of the total velocity and forces acting on the dust particle from \citep{pansp1}.}\label{fig1}
\end{figure}

\begin{figure}
  \centering {\includegraphics[width=7cm]{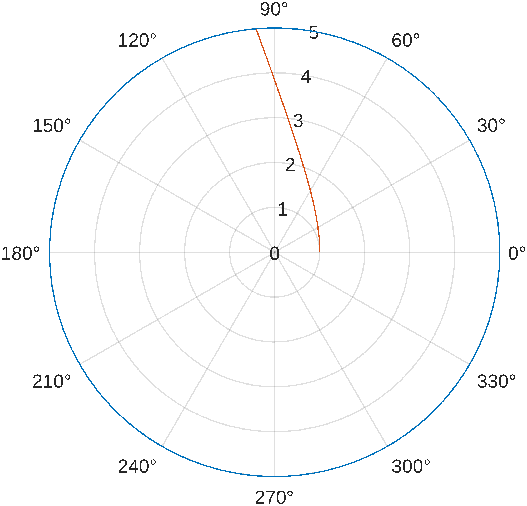}}
  \caption{Here we represent a trajectory of a dust particle (red), reaching the Jupiter's orbit (blue). The set of parameters is: $M\simeq 2\times 10^{33}$ g, $L \simeq 3.83\times 10^{33}$ erg/s, $D = 1$, $\rho = 2$ g/cm$^3$, $n_0 = 1$ cm$^{-3}$ $R(0) \simeq 1$ AU, $\dot{R}(0)\simeq 0$ km/s, and $\upsilon_{\phi}\simeq 38.3$ km/s.}\label{fig1}
\end{figure}

It is now well established that a huge amount of liquid water is localized beneath the icy surfaces of Europa \citep{europa} and Enceladus \citep{enceladus}, making the existence of life there a highly plausible possibility. Therefore, in this paper, we investigate the possibility of dust particles reaching Jupiter’s moon Europa, and consequently, the potential transport of bacteria to its surface.

The organization of the paper is the following: after outlining the mathematical model of dust particles' dynamics in Sec. 2, we obtain major results and in Sec. 3 we summarize them.

\section[]{Discussion and results}
In this section we consider dynamics of dust grains propelled from Earth reaching the nearby zone of the Jupiter's moon Europa. It is well known that dust particles with sizes of the order of $10^{-4}\; cm$ can contain bacteria packed within this size \citep{bacteria}. It is clear that, for bacteria to survive, the temperature of dust particles during their motion must not significantly exceed $T\simeq 300\; K$. Then, following an estimate presented by \cite{pansp1} the maximum velocity, $\upsilon_m$, corresponds to a regime, when a drag power, $D\rho_a\pi r^2\upsilon_m^3$ and the black body emission power $4\pi r^2\sigma T^4$ are of the same orders of magnitude
\begin{equation}
\label{power} 
v_m\simeq\left(\frac{4\sigma T^4}{\rho_a D}\right)^{1/3},
\end{equation}
where $D$ represents a drag coefficient, $\sigma$ is the Stefan-Boltzmann constant, and $\rho_a$ represents the atmospheric mass density. Following \citep{brekke,havens,picone} we consider Earth parameters as $\rho_a\simeq (1.2\times 10^{-3}\; g/cm^3)exp(-H/7.04\; km)$ , where $H$ is the altitude measured in km. By substituting the parameters at an altitude $150\; km$ with $D\simeq 1$ we obtain a velocity $\upsilon_m\simeq 14\; km/s$, which exceeds the escape velocity $\upsilon_{esc}\simeq 11.2\; km/s$. This suggests that the dust particles can overcome Earth's gravitational pull and potentially travel to other worlds. from energy conservation law one can show that at a distance of a several Earth radii, the velocity relative to Earth becomes $\upsilon_{rel}^2\simeq\upsilon_{m}^2-\upsilon_{esc}^2\simeq 8.4\; km/s$.

In the previous paper we have studied in detail the dynamics of dust particles propelled from the high altitudes of Earth. 

In Fig. 1 in polar coordinates we represent the schematic picture of velocity components and forces acting on the dust particles. Here $F_{rad}\simeq\frac{Lr^2}{4R^2c}$ is the solar radiation force \citep{carroll}, $L\simeq 3.83\times 10^{33} erg/s$ represents solar luminosity $r$ denotes the radius of a spherical dust grain, $c$ is the speed of light, $F_g = GMm/R^2$ denotes the gravitational force, $G$ is the gravitational constant, $M\simeq 2\times 10^{33}g$ represents the solar mass, $m$ is the dust particles' mass, $R$ indicates its radial coordinate, $F_{d}\approx D\rho\pi r^2\upsilon^2$ is the drag force caused by the ambient, $\upsilon$ indicates the total speed of the grain, $\rho_0\simeq 2m_pn_0$ and $n_0\simeq 1$cm$^{-3}$ are the mass density and the number density of a medium, and $m_p$ represents the proton's mass.

\cite{pansp1} has found that dynamics of dust grains can be described by the following set of equations
\begin{equation}
\label{aphi} 
\ddot{\phi}\simeq -2\dot{\phi}\frac{\dot{R}}{R}-\frac{3D\rho_0}{4\rho r}\dot{\phi}\left(\dot{R}^2+R^2\dot{\phi}^2\right)^{1/2},
\end{equation}
$$\ddot{R}\simeq R\dot{\phi}^2-\frac{3D\rho_0}{4\rho r}\dot{R}\left(\dot{R}^2+R^2\dot{\phi}^2\right)^{1/2}+$$
\begin{equation}
\label{aR} 
+
\frac{1}{R^2}\left(\frac{3L}{16\pi c\rho r}-GM\right),
\end{equation}
where $\rho$ denotes the mass density of the dust. 

In Fig. 2 we show a trajectory of the dust particles (red) moving towards the Jupiter's orbit (blue). The set of parameters is: $M\simeq 2\times 10^{33}$ g, $L \simeq 3.83\times 10^{33}$ erg/s, $D = 1$, $\rho = 2$ g/cm$^3$, $n_0 = 1$ cm$^{-3}$ $R(0) \simeq 1$ AU, $\dot{R}(0)\simeq 0$ km/s, and $\upsilon_{\phi}\simeq 38.3$ km/s. We set an initial velocity equal to $38.3 \; km/s$ where it has been taken into account that the velocity of a dust grains relative to Earth has the same direction as the Earth's velocity relative to Sun (angle, $\alpha$, between $\upsilon_{tot}$ and $\upsilon_{\phi}$ (see Fig. 1) equals zero). 

By numerically solving Eqs. (\ref{aphi},\ref{aR}) across the full range of initial values of $\alpha = 0^0$ to $360^0$, one obtains the average relative velocity of dust grains with respect to Jupiter as approximately $u_{rel}\simeq 20.1$ km/s. The impact of a dust particle on Europa’s surface results in maximum destruction when the angle of incidence — defined as the angle between the particle’s velocity vector and the surface — is $90^0$. On the other hand, it is empirically evident that $T(\theta)\simeq T(\pi/2)\sin^{3/2}\theta$ \citep{impact}. After estimating $T(\pi/2)\simeq \upsilon_{rel}^2/(2c_v)$, where $c_v\simeq 1100$ J/(kgK) is a specific heat of the dust material, and by assuming $T(\theta)\equiv T_0\simeq 300$ K, one can straightforwardly derive a critical value of the incident angle, when the particle with bacteria inside will survive
\begin{equation}
\label{tmax} 
\theta\simeq\left(\frac{2c_vT_0}{\upsilon_{rel}^2}\right)^{2/3}\simeq 0.015\;rad\simeq 1^0,
\end{equation}
Therefore, out of the entire range of possible impact angles, only a very narrow fraction — approximately $f_1 =1/360\simeq 2.8\times 10^{-3}$ — would allow for the potential survival of bacteria. These dust particles will melt Europa’s ice surface and become embedded within it. 

As discussed by \cite{pansp1,berera}, the flux density of dust particles leaving Earth is $\mathcal{F}_0\simeq 1$ cm$^{-2}$s$^{-1}$, therefore, the total flux of particles is given by $F\simeq 4\pi R_{\oplus}^2 \mathcal{F}_0\simeq 5\times 10^{18}$ s$^{-1}$, where $R_{\oplus}\simeq 6400$ km is the radius of Earth.

Numerical analysis shows that the dust grains reach the Jupiter's orbit only for the following intervals of angles $\alpha = 0^0-130^0$ $\alpha = 230^0-360^0$. Therefore, the corresponding fraction equals $f_2\simeq 260/360\simeq 0.72$. 

Another factor must be taken into account. For a particle to enter Jupiter’s zone, it must be there at the correct moment. As numerical analysis shows, particle’s angular position $\phi$ (see Fig. 1) on the Jupiter's orbit lies within the interval $83^0-135^0$, (for the case shown in Fig. 2, $\phi\simeq 95^0$) which implies a probability of approximately $f_3= (135-83)/360\simeq 0.15$ for arriving at the right time.

On average, dust particles are ejected isotropically from Earth in all directions. We now turn to estimating a maximum theoretical fraction of particles reaching a Jupiter zone. For the corresponding Hill's radius (scale of the gravitational influence), one writes $R_{_H} = r_{_J}(M_J/(3M_{\odot}))^{1/3}\simeq 0.36$ AU \citep{carroll}, where $M_J \simeq 2\times 10^{30}$ g is Jupiter's mass and $r_{_J} \simeq 5$ AU is its orbital radius. Then, one can be approximate the fraction as follows: $f_4 \simeq R_{_H}^2/(4r_{av}^2)\simeq 1.4\times 10^{-3}$. On the other hand, recent Numerical simulations have found that $0.015\%$ ($f_5 = 1.5\times 10^{-4}$) of particles entering the Jupiter's gravitation zone, can fall onto the Europa's surface.

Therefore, the total flux of dust grains on the Europa's surface with potentially survived bacteria inside is given by 
\begin{equation}
\label{totf} 
F_{surv}\simeq F\times \prod_{i=1}^{5} f_{i}\simeq 3.2\times 10^{8} \; s^{-1}.
\end{equation}
It should be noted that bacteria contained within dust particles that land on the surface undergo deactivation over a timescale of approximately $10000$ years \citep{pavlov}. On the surface of Europa there are active zones - Chaos terrains, covering approximately $20\%-40\%$ of the total surface \citep{terrain}. On the other hand, based on detailed observations, it is estimated that the surface ice undergoes fracturing for approximately $10^3 - 10^5$ years \citep{cracs}. This provides a non-negligible probability that dust particles containing microbial life may reach the subsurface water before being deactivated by radiation.

Recent studies show that the age of Europa’s icy surface is at least $30-80$ Myr \citep{age}. Therefore, the total number of particles during the mentioned period is of the order of $(3-8)\times 10^{23}$, which strongly suggests the likelihood of life being present in the subsurface ocean of Europa if the biological and biochemical conditions are compatible with Earth-originating life, which would require a new series of investigations to determine.

\section{Conclusion}
In the present paper, we have considered the dynamics of dust-grain particles propelled from Earth to Europa. For this purpose, we estimate the total flux of dust particles leaving Earth.
Considering oblique impacts, we have shown that the maximum inclination angle at which dust grains can survive is of the order of $1^\circ$. Taking into account this result, along with several important factors that reduce the flux of dust grains that impact the surface of Europe, we have found that the total flux is of the order of $3.2 \times 10^8$ s$^{-1}$. This implies that during $30-80$ Myr (age of the Europa's subsurface ocean) the total number of survived particles after impact is $(3-8)\times 10^{23}$.

We also discussed the possibility that such particles may reach the subsurface ocean through surface fracturing processes, which occur every $10^3$–$10^4$ years. This renders the existence of life on Europa highly plausible.

\section*{Acknowledgments}
The research was supported by the Shota Rustaveli National Science Foundation grant (FR-24-1751). 

\end{document}